\documentclass[pra,twocolumn,showpacs,superscriptaddress]{revtex4-1}
\usepackage{amsmath}
\usepackage{amssymb}
\usepackage{graphicx}
\usepackage{color}

\def\hafezi{\cite{hafezi}}
\def\yariv{\cite{yariv_wg}}
\def\kanemele{\cite{Kane,Z2}}

\begin{document}

\title{Optical Resonator Analog of a Two-Dimensional Topological Insulator}

\author{G.~Q.~Liang}

\affiliation{Division of Physics and Applied Physics, School of Physical and Mathematical Sciences, Nanyang Technological University, Singapore 637371, Singapore}

\author{Y.~D.~Chong}
\email{yidong@ntu.edu.sg}

\affiliation{Division of Physics and Applied Physics, School of Physical and Mathematical Sciences, Nanyang Technological University, Singapore 637371, Singapore}

\affiliation{Centre for Disruptive Photonic Technologies, School of Physical and Mathematical Sciences, Nanyang Technological University, Singapore 637371, Singapore}

\begin{abstract}
A lattice of optical ring resonators can exhibit a topological
insulator phase, with the role of spin played by the direction of
propagation of light within each ring.  Unlike the system studied by
Hafezi \textit{et al.} \cite{hafezi}, topological protection is
achieved without fine-tuning the inter-resonator couplings, which are
given the same periodicity as the underlying lattice.  The topological
insulator phase occurs for strong couplings, when the tight-binding
method is inapplicable.  Using the transfer matrix method, we derive
the bandstructure and phase diagram, and demonstrate the existence of
robust edge states.  When gain and loss are introduced, the system
functions as a diode for coupled resonator modes.
\end{abstract}


\maketitle

The idea that photonic modes can have non-trivial topological
properties, like topological states of quantum matter, originated with
Haldane and Raghu \cite{Raghu1, Raghu2}, who predicted that a
two-dimensional (2D) photonic crystal with broken time-reversal
symmetry can support modes analogous to those of a ``zero-field''
quantum Hall gas \cite{Haldane}.  This has been confirmed
experimentally, using gyromagnetic photonic crystals operating at
microwave frequencies \cite{Wang1, Wang2,Li,Li2}.  That system's most
striking feature is the existence of topologically protected one-way
photonic edge states, which could be used for on-chip isolation
\cite{Wang1}.  However, this is difficult to realize at optical
frequencies, where magneto-optic effects are weak.  Different systems
supporting topological photonic modes have subsequently been proposed
\cite{Koch,hafezi,Ochiai,Fan,Lehur,Lehur2,Caruscotto,Yannopapas,Khanikaev,Szameit}.
In particular, Hafezi \textit{et al.}~\cite{hafezi} studied a lattice
of ring resonators, similar to a 2D version of the CROW (coupled
resonator optical waveguide) \cite{CROW}, in which the direction of
propagation of light within each resonator acts as a two-fold ``spin''
degree of freedom.  In the tight-binding (weak-coupling) regime,
coupling waveguides can be used to implement spin-conserving hopping
between adjacent resonator modes, and phase shifts in these couplers
give rise to an effective vector potential in the tight-binding
hopping amplitudes, with opposite signs for the two spins.  With a
choice of phase shifts implementing the Landau gauge (which is
aperiodic in the lattice), the effective magnetic field can be made
uniform and non-zero, which yields a photonic analog of the integer
quantum Hall effect in each spin sector, with a Hofstadter butterfly
spectrum \cite{Hofstadter} and topologically protected edge states.
Although the system is reciprocal (time-reversal maps the two spin
sectors onto each other), and thus cannot be used as a conventional
optical isolator, Hafezi \textit{et al.}~suggested that the edge
states can serve as robust optical delay lines \cite{hafezi}.

The spin-dependent magnetic field in this system is reminiscent of the
topological insulator model of Kane and Mele \cite{Kane,Z2}, which has
attracted major theoretical and experimental interest \cite{Moore}.
However, there is one major difference: the couplings in the Kane-Mele
model have the periodicity of the lattice, and decoupling the two spin
sectors reduces the model to two zero-field quantum Hall systems
\cite{Haldane}, with zero net magnetic flux through each unit cell.
In the system of Hafezi \textit{et al.}, the couplings are aperiodic
and decoupled spin sectors act as \textit{integer} quantum Hall
systems; the tight-binding analysis seemed to imply that the periodic,
zero-field system is topologically trivial \cite{hafezi}.  Aperiodic
couplings also impose a practical design challenge, since a variety of
different couplers must be used.

In this Letter, we show that the zero-field resonator lattice supports
a topological insulator phase.  When the inter-resonator couplings are
tuned to large values beyond the tight-binding regime, the system
exhibits one-way edge states, with non-zero $Z_2$ topological
invariant \cite{Z2}; if the two spin sectors are decoupled, each acts
as a zero-field system, like the Kane-Mele model \cite{Kane,Z2}.  The
system therefore behaves as a photonic topological insulator.
Previously, Khanikaev \textit{et al.}~\cite{Khanikaev} have proposed a
different photonic topological insulator, which also does not require
aperiodic couplings, using linear combinations of polarization states
as the spin analog.  However, that system relies on the special
properties of metamaterials, whereas the present one uses ordinary
dielectric materials and is thus considerably more feasible.

Our calculations rely on the transfer matrix method, which has
previously been applied to the CROW \cite{yariv02, yariv_wg}, and has
a wider domain of validity for such systems than the tight-binding
method \cite{yariv_wg}.  This method also lets us easily study the
effects of gain and loss, which can produce behaviors not easily
obtainable in electronic topological insulators.  We focus on the PT
(parity/time-reversal) symmetric lattice \cite{Bender}, which contains
balanced amounts of gain and loss.  Theoretical and experimental
studies have shown that PT-symmetric lattices possess unusual
properties, including bifurcations between real and complex bands
\cite{Makris,Musslimani,Guo,Ruter,Segev,pt_nature}.  We show that in a
PT-symmetric photonic topological insulator, one edge state can be
amplified while the back-propagating state of the same spin, on the
opposite edge, is damped.  The lattice thus acts as a robust optical
diode for CROW modes.

\begin{figure}
  \centering\includegraphics[width=8.5cm]{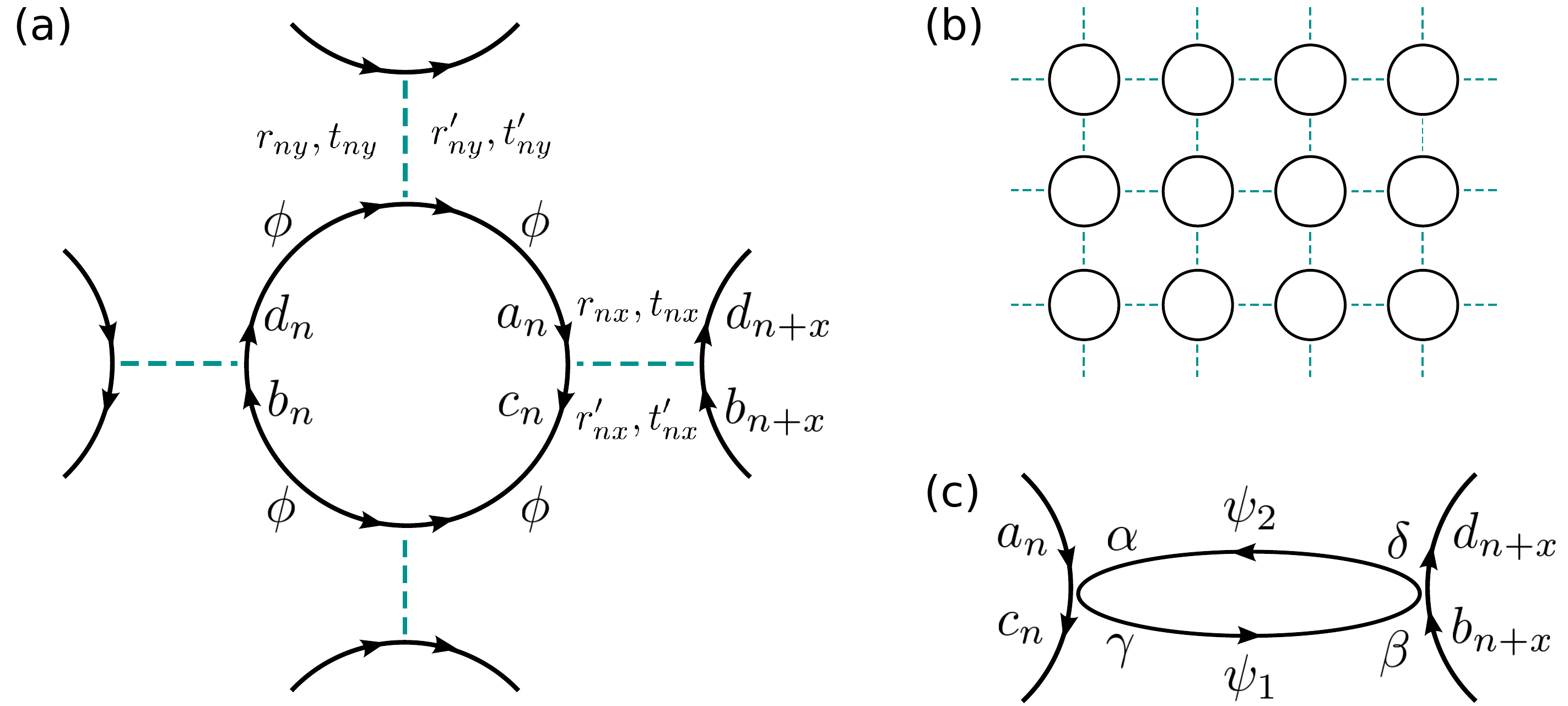}
  \caption{(color online) (a) Schematic of couplings between
    neighboring ring resonators, showing the wave amplitudes entering
    into the coupling relations (\ref{couplings}).  (b) Schematic of
    the resonator lattice over several periods.  (c) Schematic of a
    coupling waveguide which can produce the couplings shown in (a);
    $\{\alpha,\beta,\gamma,\delta\}$ label the wave amplitudes in the
    waveguides, and $\{\psi_1, \psi_2\}$ the phase shifts, which are
    used in the calculation of the coupling coefficients
    \cite{supplementary}.  }
  \label{fig:couplings}
\end{figure}

The resonator lattice is shown schematically in
Fig.~\ref{fig:couplings}.  A ring resonator occupies each site of a
square lattice.  Its modes have a two-fold ``spin'' corresponding to
the propagation direction within the ring.  As proposed in
Ref.~\cite{hafezi} and depicted in Fig.~\ref{fig:couplings}(c),
waveguides can be used to couple these modes to those on neighboring
resonators.  For our purposes, it is useful to employ a more abstract
representation for this coupling.  We first assume no spin
mixing---modes couple only to other modes of the same spin---and
restrict our attention to a single spin.  Let $n \equiv (x_n, y_n)$
denote a lattice site, $n+x$ the site one unit in the $+\hat{x}$
direction, etc.  We specify the coupling between the resonators at $n$
and $n+x$ with complex numbers $r_{nx}$, $r'_{nx}$, $t_{nx}$, and
$t'_{nx}$; similarly, we specify the coupling between $n$ and $n+y$ by
$r_{ny}$, $r'_{ny}$, $t_{ny}$, and $t'_{ny}$.  These relate the wave
amplitudes in the resonator---see
Fig.~\ref{fig:couplings}(a)---according to
\begin{equation}
      S_{nx} \begin{bmatrix} a_n \\ b_{n+x}
  \end{bmatrix} =
  \begin{bmatrix} d_{n+x} \\ c_n
  \end{bmatrix}, \;
  S_{ny} \begin{bmatrix} d_n \\ c_{n+y}
  \end{bmatrix} =
  \begin{bmatrix} b_{n+y} \\ a_n
  \end{bmatrix} \, e^{-2i\phi},
\label{couplings}
\end{equation}
where
\begin{equation}
  \quad S_{n\mu} =
  \begin{bmatrix} r_{n\mu} & t'_{n\mu} \\ t_{n\mu} & r_{n\mu}'
  \end{bmatrix}.
  \label{coupling matrices}
\end{equation}
Here, and in the following, the dummy index $\mu$ may stand for $x$ or
$y$.  The parameter $\phi$ is the phase delay across each quarter of
the ring.  The $S_{n\mu}$'s, which have the form of scattering
matrices, express the most general form of linear spin-conserving
coupling between rings.  In principle, the coefficients
$\{r_{n\mu},r_{n\mu}',t_{n\mu},t_{n\mu}'\}$ can be independently
varied by tuning the underlying waveguides \cite{supplementary}.  In
an experimental system, $\phi$ and the coupling coefficients would
depend on frequency, but here we treat them as independent quantities;
when calculating the bandstructure, $\phi$ plays the role of frequency
\cite{yariv_wg}.

Consider the special case where the coupling coefficients vary between
different sites according to
\begin{align}
  \begin{aligned}
  r_{n\mu} &= r_\mu \, e^{iA^\mu_n},&
  t_{n\mu}' &= t_\mu', \\
  t_{n\mu} &= t_\mu, &
  r_{n\mu}' &= r_\mu' \, e^{-iA^\mu_n}.
  \label{gauge field}
  \end{aligned}
\end{align}
Here, $A^x_n$ and $A^y_n$ play the role of a magnetic vector
potential.  These gauge relations generalize those used in
Ref.~\cite{hafezi}, which involved phase differences in tight-binding
hopping amplitudes.  Suppose the vector potential corresponds to a
uniform rational magnetic flux through each unit cell: $A^x_n +
A^y_{n+x} - A^x_{n+y} - A^y_{n} = 2\pi P/Q$, where $P$ and $Q$ are
integers.  For $Q = 1$, i.e.~integer flux through each unit cell, the
bandstructure is the same as in the zero-field ($A^x_n = A^y_n = 0$)
system.  Then the magnetic unit cell coincides with the lattice's unit
cell, and there are solutions of the form \cite{Hofstadter,yariv_wg}
\begin{equation}
  a_{n+\mu} = e^{i(K_\mu + A_n^\mu)} a_n, \;\quad
  b_{n+\mu} = e^{i(K_\mu + A_n^\mu)} b_n, \label{magnetic Bloch}
\end{equation}
where $K_\mu$ is a Bloch wave-vector.  Combining
(\ref{couplings})-(\ref{magnetic Bloch}) gives \cite{supplementary}:
\begin{align}
  \begin{aligned}
  &e^{-4i\phi} - B e^{-2i\phi} - C = 0, \\
  &B = r_x't_y' e^{iK_x} + r_xt_y e^{-iK_x}
  + t_x r_y' e^{iK_y} + t_x' r_y e^{-iK_y} \\
  &C = (r_xr_x'-t_xt_x') (r_yr_y' - t_yt_y').
  \label{dispersion}
  \end{aligned}
\end{align}
As we shall see, for unitary couplings this gives rise to four real
bands in the periodic space $\phi \in [-\pi,\pi]$: two in
$[-\pi/2,\pi/2]$ from directly solving (\ref{dispersion}), and the
other two by adding $\pm \pi$.  This result relies crucially on the
fact that in Eq.~(\ref{gauge field}) there is no phase variation in
$t_{n\mu}$ and $t_{n\mu}$.  The coupler shown in
Fig.~\ref{fig:couplings}(c) satisfies this condition if the sum of the
phase delays on its two arms is kept constant \cite{supplementary}.
For non-integer fluxes ($Q \ne 1$), the current approach gives
essentially the same results as Ref.~\cite{hafezi}: we could impose
the Landau gauge $A^x_n = (P/Q)\, y_n$ and $A^y_n = 0$, and define a
$Q \times 1$ magnetic unit cell for which $a_{n+Qx} = e^{i(K_x + P
  y_n)} a_n$ and $a_{n+y} = e^{iK_y} a_n$, and similarly for $b$.
This gives $4Q$ bands, analogous to Landau levels.

In the remainder of this paper, we focus on the zero-field (integer
flux) system.  If the couplings conserve energy, then $S^\dagger_\mu =
S^{-1}_\mu$.  We expect the bandstructure $\phi(K_x,K_y)$ to be real
(for $K_x$ and $K_y$ real), and this is easily proven using the
parameterization
\begin{align}
  \begin{aligned}
    r_\mu &= \sin\theta_\mu \, e^{i\chi_\mu}, &
    t_\mu' &= - \cos\theta_\mu \, e^{i(\varphi_\mu - \xi_\mu)}, \\
    t_\mu &= \cos\theta_\mu \, e^{i\xi_\mu}, &
    r_\mu' &= \sin\theta_\mu \, e^{i(\varphi_\mu - \chi_\mu)},
    \label{unitary parameters}
  \end{aligned}
\end{align}
where $\theta_\mu \in [0, \pi/2]$ and $\chi_\mu, \xi_\mu, \varphi_\mu
\in [0, 2\pi]$.  Eq.~(\ref{dispersion}) then simplifies to
$e^{-4i\tilde{\phi}} + 2i\, Y \, e^{-2i\tilde{\phi}} - 1 = 0$, where
\begin{align}
  \begin{aligned}
  Y &\equiv \sin\theta_x \cos\theta_y \sin \tilde{K}_x - \cos\theta_x \sin\theta_y \sin\tilde{K}_y \\
  \tilde{\phi} &\equiv \phi + \frac{\varphi_x+\varphi_y}{4}\\
  \tilde{K}_x &\equiv K_x + \frac{\varphi_x}{2} - \chi_x + \frac{\varphi_y}{2} - \xi_y \\
  \tilde{K}_y &\equiv K_y + \frac{\varphi_y}{2} - \chi_y - \frac{\varphi_x}{2} + \xi_x.
  \label{simple dispersion}
  \end{aligned}
\end{align}
For real $K_\mu$, $\left|Y\right| \le \sin(\theta_x + \theta_y) \; \le
\; 1$, and the bands are
\begin{align}
  \begin{aligned}
  \phi_+ &= m\pi - \frac{\varphi_x + \varphi_y}{4}
  + \frac{1}{2} \sin^{-1} \left[Y(K_\mu)\right] \\
  \phi_- &= n\pi\, - \frac{\varphi_x + \varphi_y}{4}
  + \frac{1}{2}\left\{\pi - \sin^{-1} \left[Y(K_\mu)\right]\right\}.
  \end{aligned}
\end{align}

\begin{figure}
  \centering\includegraphics[width=8.5cm]{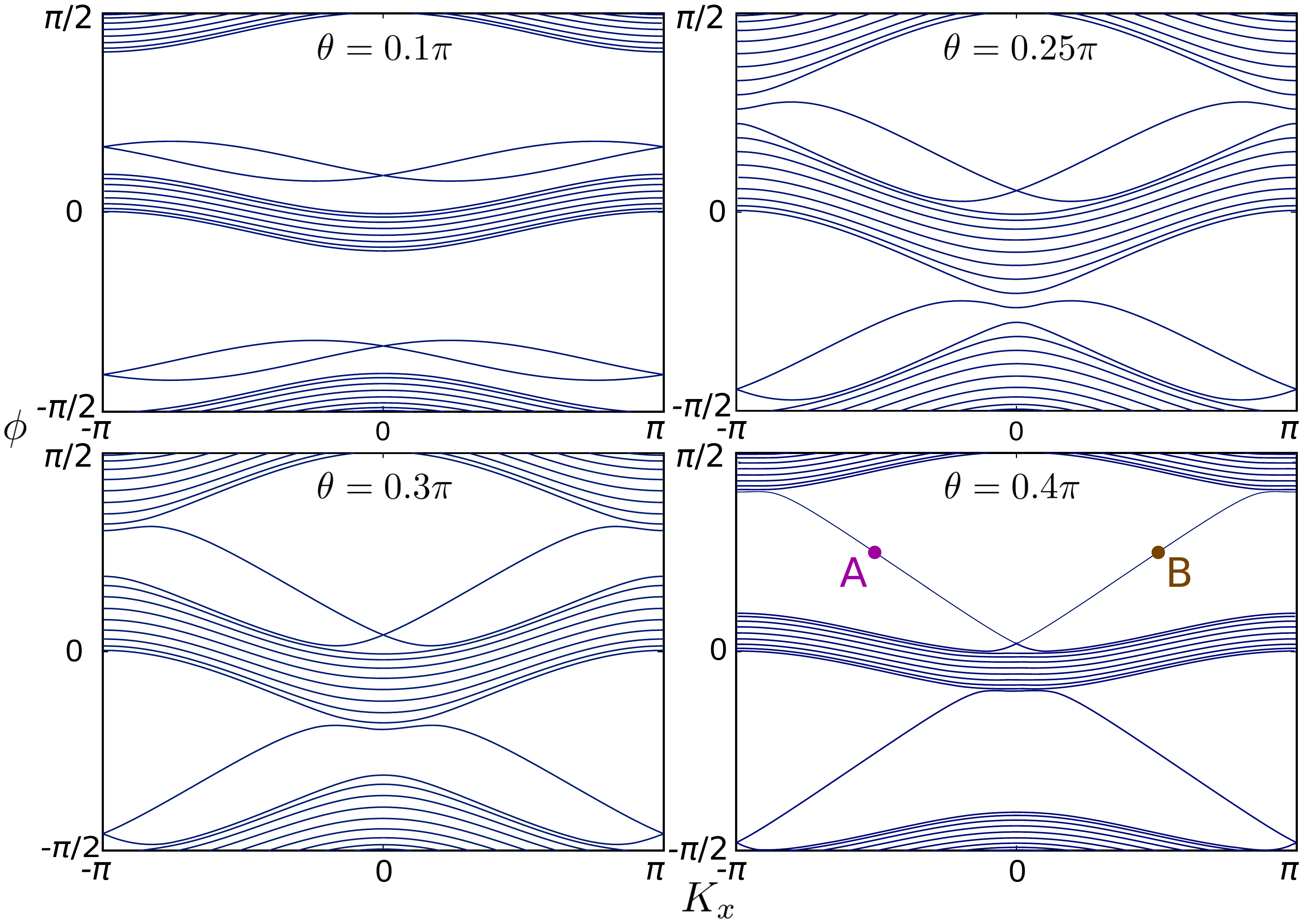}
  \caption{(color online) Projected band diagram of a semi-infinite
    resonator lattice, with 10 cells in the $y$ direction.  The spin
    sectors are decoupled; the model parameters are given by
    Eq.~(\ref{unitary parameters}) with $\varphi_\mu = \chi_\mu = 0$,
    $\xi_\mu = \pi/2$, and $\theta_x = \theta_y = \theta$.  Band
    crossing occurs at $\theta = \pi/4$, and the system is a
    topological insulator for $\theta > \pi/4$.  For $\theta =
    0.4\pi$, the points labeled $A$ and $B$, at $\phi = \pi/4$,
    indicate the edge states plotted in Fig.~\ref{fig:edgeintensity}.}
  \label{fig:projected}
\end{figure}

\begin{figure}
  \centering\includegraphics[width=8.5cm]{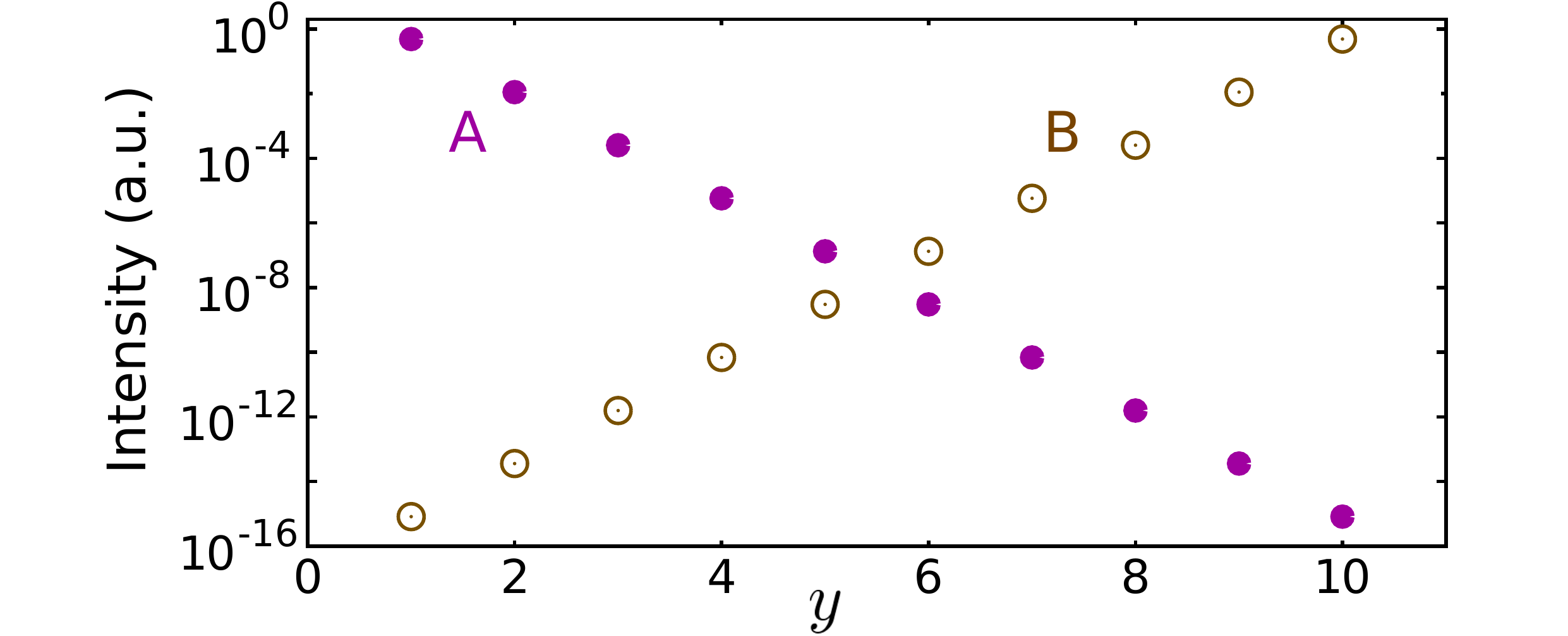}
  \caption{(color online) Semi-log plot of edge state intensity versus
    $y$ lattice coordinate, demonstrating edge confinement.  The edge
    states, labeled $A$ (filled circles) and $B$ (open circles), have
    equal $\phi = \pi/4$, and occur at $K_x = \mp 1.587$ respectively.
    The parameters are the same as in Fig.~\ref{fig:projected}, with
    $\theta = 0.4\pi$.  The spins are clockwise, as depicted in
    Fig.~\ref{fig:couplings}.  The intensities are defined as the
    value of $(|a_n|^2 + |b_n|^2 + |c_n|^2 + |d_n|^2)/4$ in each
    resonator.}
  \label{fig:edgeintensity}
\end{figure}

The above calculation also yields the phase diagram.  Band-crossing
points occur where the inequality saturates: $\theta_x + \theta_y =
\pi/2$, or equivalently $|r_x|^2 + |r_y|^2 = 1$.  This defines a
boundary between two insulator phases.  To show that one of these
phases is topologically non-trivial, we specialize to $\varphi_\mu =
\chi_\mu = 0$, $\xi_\mu = \pi/2$, and $\theta_x = \theta_y = \theta$,
so that
\begin{equation}
  Y = - \frac{1}{2} \sin2\theta \; \left[\cos K_x + \cos K_y\right].
  \label{simpley}
\end{equation}
The projected band diagram for a semi-infinite strip can be calculated
similarly \cite{supplementary}, with results shown in
Fig.~\ref{fig:projected}.  For $\theta < \pi / 4$, the system is a
trivial insulator; although Fig.~\ref{fig:projected}(a) exhibits edge
states for some $\phi$, these are two-way edge states, and for each
$\phi$ there are states confined to the same edge at different $K_x$,
with positive as well as negative group velocities.  For $\theta >
\pi/4$, the system is a topological insulator.  The edge states span
the band gaps, and for the given spin (clockwise) there is a positive
velocity upper edge state and a negative velocity lower edge state
(Fig.~\ref{fig:edgeintensity}).  In a real system, where the model
parameters depend on the frequency $\omega$, the topologically
non-trivial band gaps would correspond to frequencies for which
$\theta(\omega) > \pi /4$.  We have verified, using finite-difference
time-domain simulations, that this strong-coupling regime can be
achieved with realistic resonator and waveguide designs
\cite{supplementary}.

It is noteworthy that the topological insulator phase occurs only when
the inter-resonator coupling is sufficiently strong.  This phase does
not appear in the tight-binding analysis, where the zero-field system
appears to be topologically trivial \cite{hafezi}.  The transfer
matrix method, however, accounts for the wave amplitudes at different
parts of each ring, which is needed to describe the edge states of the
topological insulator phase.  Roughly speaking, these edge states move
in the same direction in which light propagates inside the upper
(lower) half of the uppermost (lowermost) ring resonator of the strip.

Spin mixing can be induced by backscattering within the resonators or
waveguides \cite{hafezi}.  This lifts the spin degeneracy of the edge
states \cite{supplementary}, similar to the Rashba term in electronic
topological insulators \cite{Kane,Z2}.  If the couplings remain
unitary and reciprocal (i.e.,~absent radiative loss and magneto-optic
disorder), the states on each edge are Kramers pairs
\cite{supplementary}.  However, these edge states are not
topologically protected against spin-mixing perturbations, because
optical wave amplitudes, unlike electrons, are not spin-half objects
\cite{Kane}.

\begin{figure}
  \centering\includegraphics[width=8.6cm]{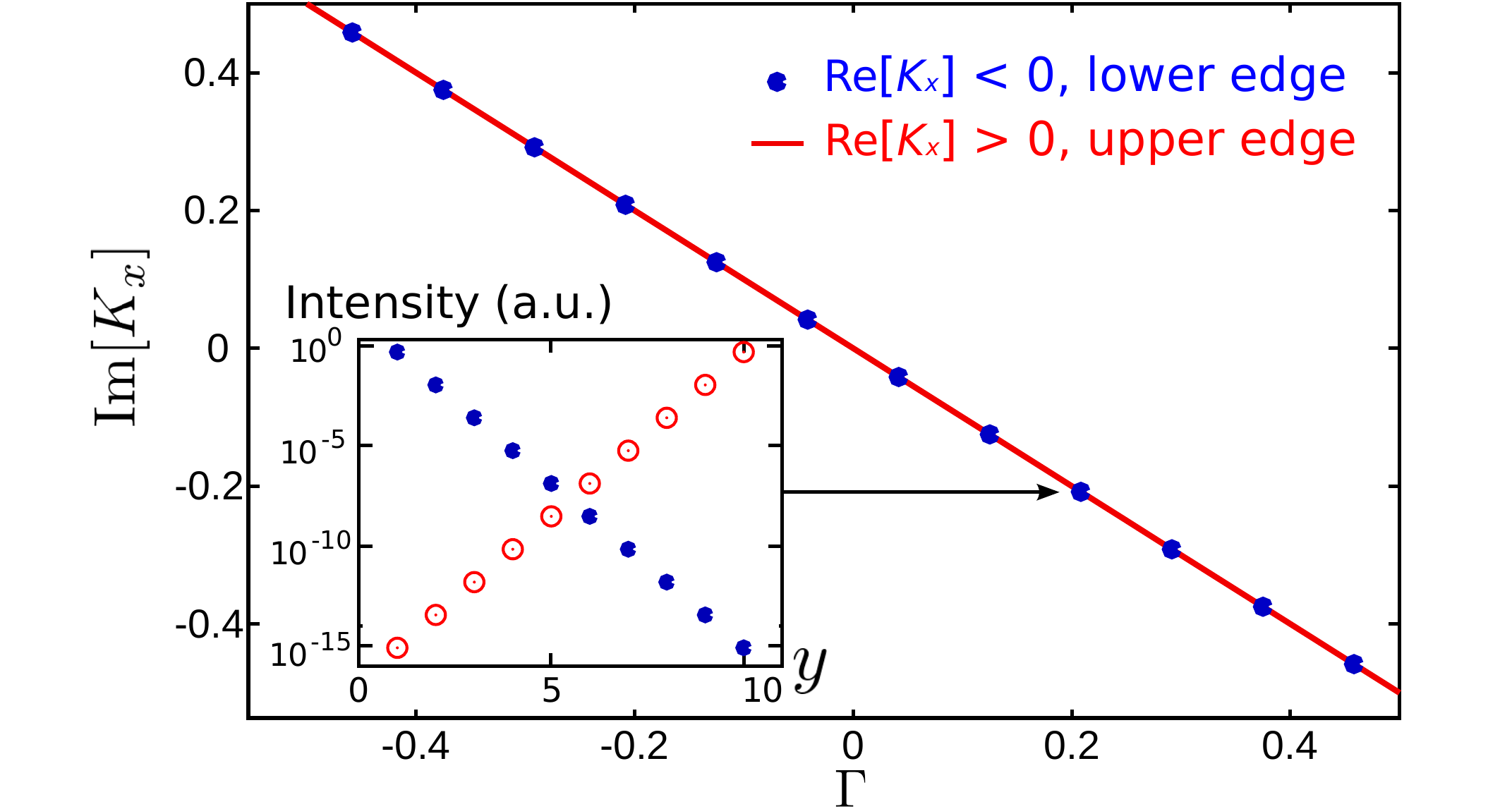}
  \caption{(color online) Amplification and damping of edge states in
    the PT-symmetric resonator lattice.  $\Gamma$ is the gain-loss
    parameter, defined in Eq.~(\ref{pt parameterization}).  All other
    parameters are the same as in Fig.~\ref{fig:projected}, with
    $\theta = 0.4\pi$.  Both edge states acquire the same value of
    $\textrm{Im}[K_x]$, so one is damped and the other amplified.
    Inset: intensity profiles for the lower edge state (filled
    circles) and upper edge state (open circles) at $\Gamma = 0.2$.  }
  \label{fig:pt}
\end{figure}

\begin{figure}
  \centering\includegraphics[width=8.6cm]{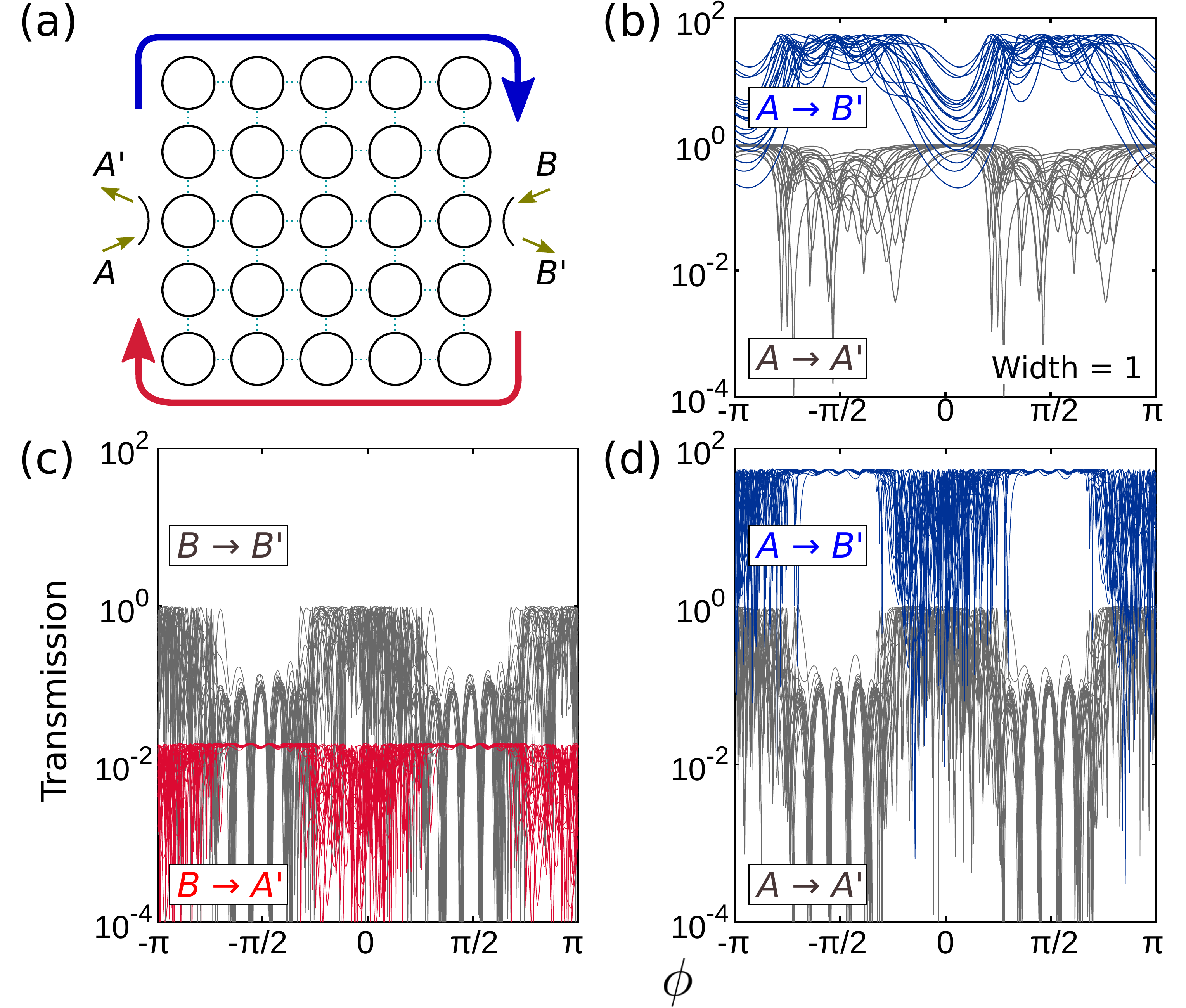}
  \caption{Transmittance across a disordered PT-symmetric resonator
    lattice.  (a) Schematic.  (b) Transmittance from port $A$ to $B'$
    (blue) for a one unit cell wide lattice, which has no topological
    protection.  (c,d) Transmittance leftward from $B$ to $A'$ (red),
    and rightward from $A$ to $B'$ (blue), when the lattice is 5 unit
    cells wide as shown in (a).  Reflectances are shown in grey.  In
    (b)-(d), the lattice is 5 cells long, and transmittances are
    plotted for 20 disorder realizations, where each coupling has
    random $\theta$, distributed uniformly in $[0.2\pi,0.5\pi]$.  The
    $x$-couplings have $\Gamma = 0.5$. }
  \label{fig:diode}
\end{figure}

We have studied the effects of incorporating gain and loss into the
photonic topological insulator, which yields behaviors that are
inaccessible in the electronic system \cite{Makris}.  In particular,
we consider here the PT-symmetric case, which corresponds to putting
``balanced'' gain and loss in symmetric regions of the unit cell.
PT-symmetric photonic systems have previously been studied
experimentally, e.g.~using lossy waveguides \cite{Guo} and optical
fiber systems \cite{pt_nature}.  In the present system, the transfer
matrix method can be adapted to include gain and loss simply by making
the coupling matrices non-unitary.  Specifically, the matrices obey
the PT-symmetry relation \cite{Schomerus,ChongPT,LiPT}
\begin{equation}
  \mathcal{PT}\, S_\mu \mathcal{PT} = S^{-1}_\mu.
  \label{PT symmetry}
\end{equation}
Here, $\mathcal{P}$ and $\mathcal{T}$ are parity and time-reversal
operators.  We choose $\mathcal{P} = [0,1;1,0]$ and $\mathcal{T}$ to
be the complex conjugation operator; for the coupler shown in
Fig.~\ref{fig:couplings}(c), setting $\psi_2 = \psi_1^*$ satisfies
Eq.~(\ref{PT symmetry}).  The $S_\mu$'s can then be parameterized by
$r = |r|\, e^{i\varphi}$, $t' = -|t|\, e^{i(\varphi-\varphi')}$, $t =
|t|\, e^{i(\varphi+\varphi')}$, and $r' = |r'|\, e^{i\varphi}$, where
$|rr'| + |t|^2 = 1$ \cite{PT note}. For simplicity, we set $\varphi =
0$ and $\varphi' = \pi/2$, so that
\begin{align}
  \begin{aligned}
    r &= e^\Gamma \sin\theta, & t' &= i\cos\theta, \\
    t &= i\cos\theta, &r' &= e^{-\Gamma} \sin\theta,
    \label{pt parameterization}
  \end{aligned}
\end{align}
where $\Gamma$ characterizes the amount of gain and loss.

Fig.~\ref{fig:pt} shows the effects of PT-symmetric gain and loss on
the edge states of the photonic topological insulator.  We assume no
spin mixing; $\Gamma$ is varied for the $x$ couplings, while the $y$
couplings are kept unitary ($\Gamma = 0$) \cite{note1}.  For the bulk
bands, Eq.~(\ref{pt parameterization}) causes $K_x$ to be replaced by
$K_x - i \Gamma$ in the solution (\ref{simpley}), so that the bands
are real for $K_x = m \pi$, $m\in \mathbb{Z}$, and complex otherwise.
The edge states on opposite edges of the semi-infinite strip, which
have opposite velocities, acquire the same imaginary component to
$K_x$, and are respectively amplified and damped.  This has a simple
interpretation.  The upper edge state's wave amplitude is multiplied
by $r_x$ each time it hops one ring to the right; for $\Gamma > 0$,
$|r_x| > 0$ and hence the state is amplified.  Likewise, the lower
edge state is damped by $r_x'$ with each leftward hop.  Previous
studies of PT-symmetric waveguides have shown that modes with
different transverse profiles can be selectively amplified and damped
\cite{Guo}, but in those waveguides each amplified (damped) mode has a
counter-propagating partner which is amplified (damped) by an equal
amount.  Here, the edge states have no counter-propagating partners of
the same spin.

Fig.~\ref{fig:diode} shows the transmittance between waveguides
coupled to opposite ends of the finite PT-symmetric lattice.
Left-to-right transmission is amplified, while transmission in the
opposite direction is damped.  Within the band gaps, the transmission
is insensitive to disorder, due to the topological protection on the
edge states.  In Fig.~\ref{fig:diode}(b), we test the effect of
removing this topological protection by performing the calculation
with the lattice width reduced to a single unit cell; the resulting
transmission is considerably less stable, varying by an order of
magnitude for the same values of $\phi$ \cite{supplementary}.  In
terms of the underlying waveguides, the system is reciprocal, but it
can nonetheless serve as a diode element for CROW modes.  Such modes
are susceptible to backscattering, even in the absence of
spin mixing \cite{hafezi}; this is a particular problem in
slow-light applications \cite{CROW backscattering}.  A photonic
topological insulator can offset the effects of backscattering loss by
robustly amplifying forward modes and damping backward modes.  Unlike
the PT-symmetric diode of Ref.~\cite{diode}, this device does not
require optical nonlinearity.

This research was supported by the Singapore National Research
Foundation under grant No.~NRFF2012-02.  We thank J.~M.~Taylor and
M.~Hafezi for helpful discussions.

\newpage

\begin{widetext}

\appendix
\section{Inter-loop Coupling Coefficients}

In Eq.~(1) of the main text, we abstract away the couplings between
resonators into a set of ``reflection coefficients'' and
``transmission coefficients'', reproduced here for convenience:
\begin{align}
  S_{nx} & \begin{bmatrix} a_n \\ b_{n+x}
  \end{bmatrix} =
  \begin{bmatrix} d_{n+x} \\ c_n
  \end{bmatrix},   \label{coupling 1} \\
  S_{ny} & \begin{bmatrix} d_n \\ c_{n+y}
  \end{bmatrix} =
  \begin{bmatrix} b_{n+y} \\ a_n
  \end{bmatrix} \, e^{-2i\phi}, \label{coupling 2}\\
  \quad S_{n\mu} &=
  \begin{bmatrix} r_{n\mu} & t'_{n\mu} \\ t_{n\mu} & r_{n\mu}'
  \end{bmatrix}.
  \label{coupling 3}
\end{align}
In this section, we discuss how these coefficients can be related to
the parameters of the underlying waveguide-coupling mechanism, such as
that discussed in Ref.~\hafezi.  As shown in Fig.~1(c), the
resonator amplitudes $a$, $b$, $c$, and $d$ (we will omit the
redundant subscripts for simplicity) are coupled to waveguide
amplitudes $\alpha$, $\beta$, $\gamma$, and $\delta$, by coupling
relations \yariv:
\begin{eqnarray}
  \begin{bmatrix}\tau & i\kappa \\ i\kappa & \tau
  \end{bmatrix}  \begin{bmatrix} a \\ \alpha
  \end{bmatrix} &=& \begin{bmatrix} c \\ \gamma \end{bmatrix}
  \label{eqa1} \\
  \begin{bmatrix}\tau' & i\kappa' \\ i\kappa' & \tau'
  \end{bmatrix}  \begin{bmatrix} b \\ \beta
  \end{bmatrix} &=& \begin{bmatrix} d \\ \delta \end{bmatrix}.
\end{eqnarray}
Without loss of generality, $\tau$, $\tau'$, $\kappa$, and $\kappa'$
may be taken to be real.  If the phase delays along the two arms of
the waveguide are $\psi_1$ and $\psi_2$ respectively, then
\begin{equation}
  \alpha = e^{i\psi_2}\, \delta, \quad \beta = e^{i\psi_1}\, \gamma.
  \label{eqa3}
\end{equation}
By combining Eqs.~(\ref{eqa1})-(\ref{eqa3}), we obtain
\begin{equation}
  \begin{bmatrix} d \\ c \end{bmatrix}
  = \frac{1}{1-\tau\tau'e^{i(\psi_1+\psi_2)}}\,
  \begin{bmatrix} -\kappa\,\kappa'\, e^{i\psi_1} &
    \tau'-\tau\,e^{i(\psi_1+\psi_2)} \\
    \tau-\tau' \, e^{i(\psi_1+\psi_2)} & -\kappa \kappa' e^{i\psi_2}
  \end{bmatrix}
  \begin{bmatrix} a \\ b \end{bmatrix}.
  \label{effective matrix}
\end{equation}
Comparison with (\ref{coupling 1})-(\ref{coupling 3}) immediately
yields the corresponding values of $r$, $r'$, $t$, and $t'$.

As indicated in Ref.~\hafezi, a synthetic magnetic vector potential
can be implemented by altering the phase delay $\psi_1$ at different
sites, while keeping $\psi_1 +\psi_2$ fixed.  This leaves $t$ and
$t'$, which are given by the off-diagonal matrix elements in
(\ref{effective matrix}), independent of $\psi_1$, in agreement with
Eq.~(3).  The relevant component of the resulting vector potential is
$A = \psi_1$.

\begin{figure}
  \centering\includegraphics[width=12cm]{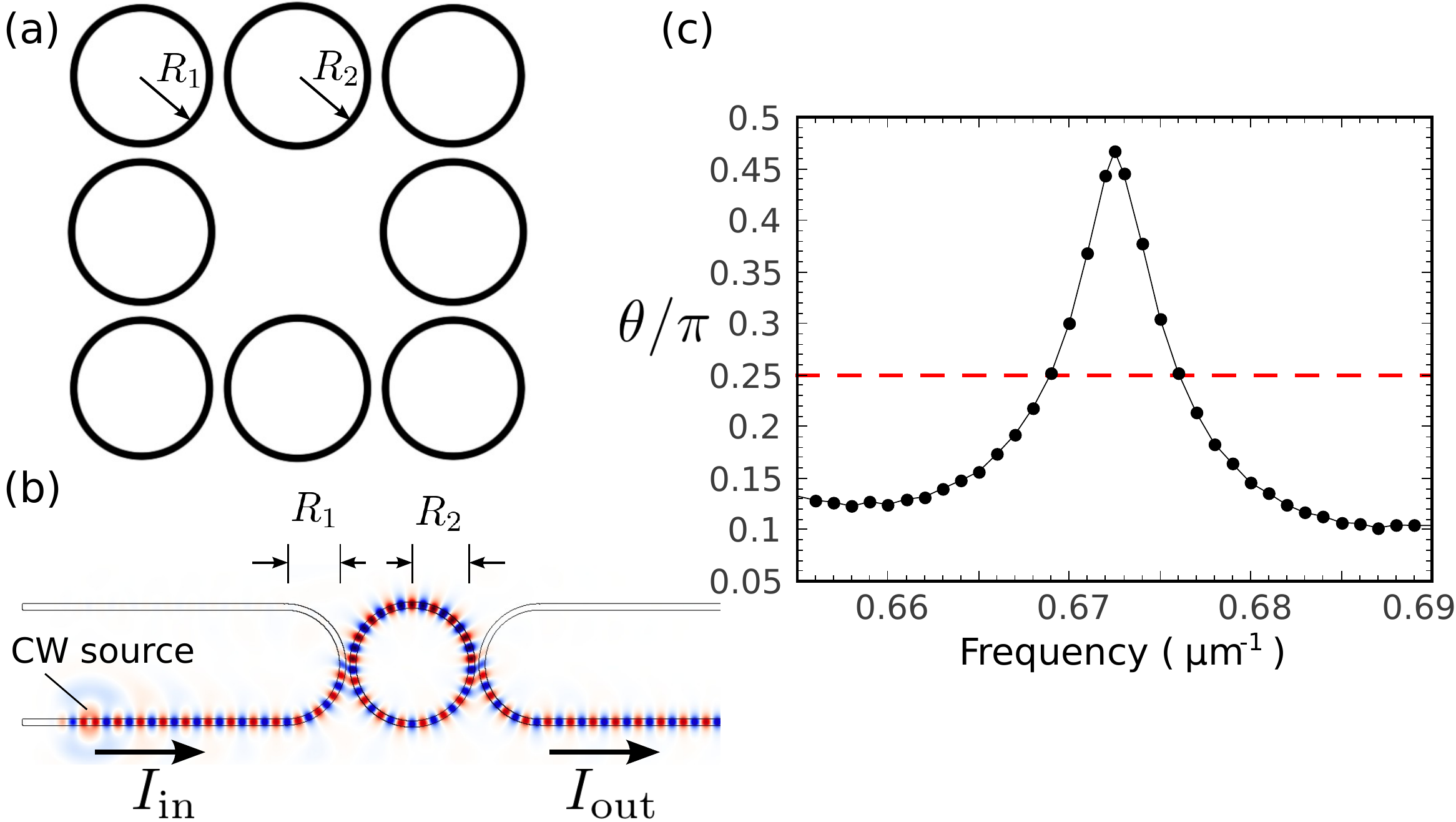}
  \caption{(a) Realization of the resonator lattice system, using ring
    resonators as couplers.  (b) Simulation cell for calculating the
    coupling parameter $\theta$.  (c) Plot of $\theta$ versus
    frequency.}
  \label{fig:theta}
\end{figure}

\begin{figure}
  \centering\includegraphics[width=8cm]{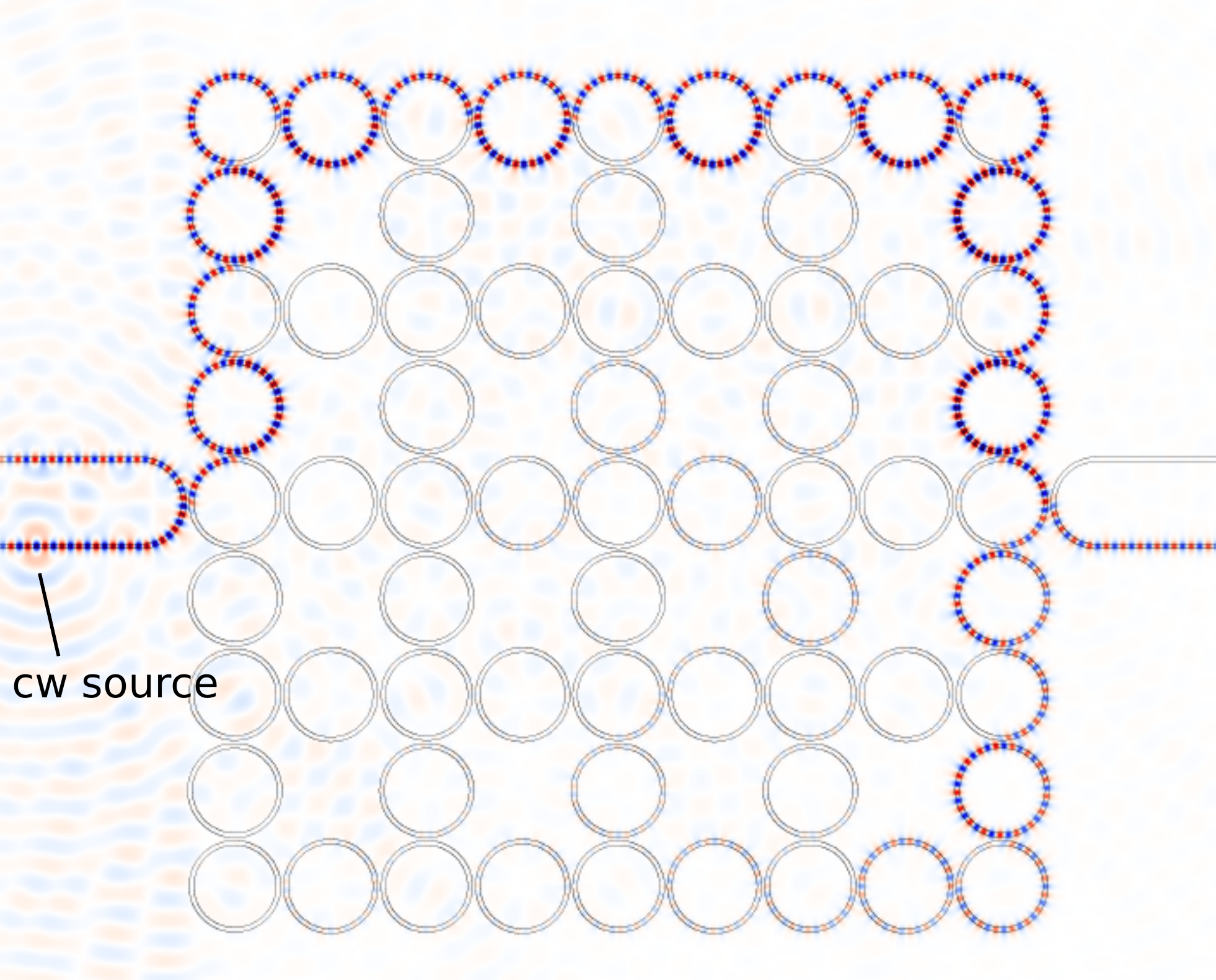}
  \caption{FDTD simulation of a photonic topological insulator
    lattice.  All lattice parameters are the same as in
    Fig.~\ref{fig:theta}, and the operating frequency is 0.6725
    $\mu\textrm{m}^{-1}$, in the middle of the bandgap indicated by
    Fig.~\ref{fig:theta}(c).  An edge state is observed, as predicted
    by the transfer matrix theory. }
  \label{fig:fdtd}
\end{figure}

A realization of this coupling scheme, using optical ring resonators,
is shown in Fig.~\ref{fig:theta}(a).  Each ring has width 200 nm and
refractive index $n = 3$ (with $n = 1$ in the surrounding space).  The
rings at lattice sites have inner radius $R_1 = 1.6$ $\mu$m, and the
rings serving as couplers have inner radius $R_2 = 1.653$ $\mu$m.  The
gap between each ring is fixed at 100\,nm.  As discussed in the main
text, the coupling parameters entering into the $S$ matrices depend
implicitly on the operating frequency.  Using finite-difference
time-domain (FDTD) simulations of the structure shown in
Fig.~\ref{fig:theta}(b), we extract the parameter $\theta$ as a
function of frequency, via the relation $\theta =
\sin^{-1}(\sqrt{I_{\textrm{out}}/I_{\textrm{in}}})$ where
$I_{\textrm{in}}$ and $I_{\textrm{out}}$ are the intensities at the
input and output ports.  The result is shown in
Fig.~\ref{fig:theta}(c).  The strong-coupling regime $\theta > \pi/4$
is found to be achievable, e.g.~within the frequency range
0.669--0.676 $\mu\textrm{m}^{-1}$.  In Fig.~\ref{fig:fdtd}, we show
the results of an FDTD simulation of a finite lattice at frequency
0.6725\,$\mu\textrm{m}^{-1}$, demonstrating that the system indeed
behaves as a photonic topological insulator.  We thus conclude that
on-chip realizations of the photonic topological insulator are quite
feasible.  More detailed simulation studies of such systems will be
presented in a subsequent paper.

\section{Gauge Transformations}

In this section, we provide additional details about the gauge
structure of the resonator lattice.  The coupling relations between
resonator amplitudes, given by (\ref{coupling 1})-(\ref{coupling 3}),
can be simplified by eliminating the $c$ and $d$ variables, which
yields
\begin{align}
  \begin{aligned}
  e^{-2i\phi} b_{n} - r_{{n-y},y} \, t_{n-x-y,x}' \, b_{n-y}
  - t_{n-y,y}' \, r_{n,x}' \, b_{n+x}
  &=
  r_{n-y,y} \, r_{n-x-y,x} \, a_{n-x-y}
  + t_{n-y,y}' \, t_{n,x} \, a_n \\
  e^{-2i\phi} a_{n} - t_{ny} \, r_{n-x,x} \, a_{n-x}
  - r_{n,y}' \, t_{n+y,x} \, a_{n+y}
  &= r_{ny}' \, r_{n+y,x}' \, b_{n+x+y} + t_{n,y} \, t_{n-x,x}' \, b_n.
  \label{ab1}
  \end{aligned}
\end{align}
When the coupling coefficients obey the gauge relations (3), the
system of equations (\ref{ab1}) simplifies to
\begin{align}
  \begin{aligned}
  e^{-2i\phi} b_{n} - t_x' r_y e^{iA^y_{n-y}} b_{n-y}
  - r_x' t_y' e^{-iA^x_n} b_{n+x} &=
  r_x r_y e^{i(A^x_{n-x-y} + A^y_{n-y})} a_{n-x-y}
  + t_x t_y' a_n \\
  e^{-2i\phi} a_{n} - r_x t_y e^{iA^x_{n-x}} a_{n-x}
  - t_x r_y' e^{-iA^y_n} a_{n+y}
  &= r_x' r_y' e^{-i(A^x_{n+y} + A^y_n)} b_{n+x+y}
  + t_x' t_y b_n.
  \label{ab2}
  \end{aligned}
\end{align}
We look for solutions to these two equations, using the Bloch ansatz
(4).  When writing down the amplitudes for negative displacements
and/or composed displacements using this ansatz, some care is needed
to ensure that the $A$'s are evaluated at lattice positions consistent
with (4).  For example,
\begin{align}
  \begin{aligned}
  a_{n-x} &= e^{-i(K_x+A_{n-x}^x)} a_n \\ a_{n+x+y} &= e^{i(K_x+K_y
    + A_{n+y}^x + A_n^y)} a_n \\ a_{n-x-y} &=
  e^{-i(K_x+K_y+A_{n-x-y}^x+A_{n-y}^y)} a_n, \;\; \textrm{etc.}
  \end{aligned}
\end{align}
Plugging these relations into (\ref{ab2}) yields
\begin{align}
  \begin{aligned}
    \Big[e^{-2i\phi} - r_x' \, t_y' \,e^{iK_x}
      - t_x' \, r_y \, e^{-iK_y} \Big] \, b_n &=
    \Big[ r_x \, r_y \, e^{-i(K_x+K_y)} + t_x \, t_y' \Big] \, a_n \\
    \Big[e^{-2i\phi} - r_x \, t_y \, e^{-iK_x}
      - t_x \, r_y' \, e^{iK_y} \Big] \, a_n
    &= \Big[r_x' \, r_y' \, e^{i(K_x+K_y)} + t_x' \, t_y \Big] \, b_n.
    \label{ab3}
  \end{aligned}
\end{align}
The resulting dispersion relation ($\phi$ versus $K_x$ and $K_y$) is
independent of $A^x_n$ and $A^y_n$.

\section{Projected band structure}

The calculation of the projected band diagram for the resonator
lattice is much like the calculation of the CROW's dispersion relation
by Yariv \textit{et al.}~\yariv.  Suppose we have a semi-infinite
strip, whose width is $N$ lattice sites in the $y$ direction and
infinite in the $x$ direction, with the coupling matrices independent
of the $x$ coordinate.  First, assume as before that the spins are
decoupled and consider a single spin sector (clockwise).  We seek
solutions of the form
\begin{equation*}
  [a,b,c,d]_{n+x} = e^{iK_x} [a,b,c,d]_n.
\end{equation*}
Let us consider a single column of the lattice, and replace the site
index $n$ with $j \in \{1,2,\cdots,N\}$ denoting the $y$ coordinate.
From (\ref{coupling 1}),
\begin{equation}
  M_j^x \begin{bmatrix}c_j \\ a_j \end{bmatrix}
  = e^{iK_x} \, \begin{bmatrix} b_j \\ d_j  \end{bmatrix}
  \quad \textrm{where}\;\;
  M_j^x \equiv \frac{1}{r_{jx}'} \begin{bmatrix}
    1 & -t_{jx} \\ t_{jx}' & \det[S_{jx}] \end{bmatrix}.
  \label{Mj}
\end{equation}
From (\ref{coupling 2}),
\begin{equation}
  M^y_j(\phi) \begin{bmatrix}d_j \\ b_{j+1} \end{bmatrix}
  = \begin{bmatrix} a_j \\ c_{j+1}  \end{bmatrix}
  \quad \textrm{where}\;\;
  M^y_j(\phi) \equiv \frac{1}{t_{jy}'} \begin{bmatrix}
    -e^{2i\phi} \det[S_{jy}] & r_{jy}' \\
    -r_{jy} & e^{-2i\phi} \end{bmatrix},
  \label{Mp}
\end{equation}
for $j = 1, \cdots, N-1$.  At the edges of the lattice, we have the
boundary conditions (refer to Fig.~1 of the paper):
\begin{equation}
  c_1 = e^{-2i\phi} b_1, \quad a_N = e^{2i\phi} d_N.
  \label{boundary}
\end{equation}
Combining (\ref{Mj})-(\ref{boundary}) yields the eigenvalue equation
\begin{equation}
  M_{A} M_{B} \begin{bmatrix} b_1 \\ d_1 \\ \vdots \\ b_N \\ d_N \end{bmatrix}
  = e^{iK_x} \begin{bmatrix} b_1 \\ d_1 \\ \vdots \\ b_N \\ d_N \end{bmatrix},
  \label{eigenproblem}
\end{equation}
where $M_A$ and $M_B$ are the following $2N\times 2N$ matrices:
\begin{equation}
  M_A = \begin{bmatrix} M_1^x & & \\
    & M_2^x & & \\
    & & \ddots & \\
    & & & M_N^x
  \end{bmatrix}, \quad
  M_B = \begin{bmatrix} e^{-2i\phi} & & & & \\
    & M_1^y & & & \\
    & & \ddots & & \\
    & & & M_{N-1}^y & \\
    & & & & e^{2i\phi}
  \end{bmatrix}.
\end{equation}
To generate the projected band diagram, we sweep $\phi$ through the
desired range (usually $[-\pi/2, \pi/2]$ due to the $\pi$-periodicity
of the band structure), solving the eigenvalue problem
(\ref{eigenproblem}) at each $\phi$.  The eigenvalues which do not
have unit modulus, which correspond to evanescent modes, are
discarded; the arguments of the rest are the value(s) of $K_x$ for
that value of $\phi$.  This yields the plots shown in Fig.~2 of the
paper.

\section{Transfer matrix analysis for both spins}

\begin{figure}[b]
  \centering\includegraphics[width=4cm]{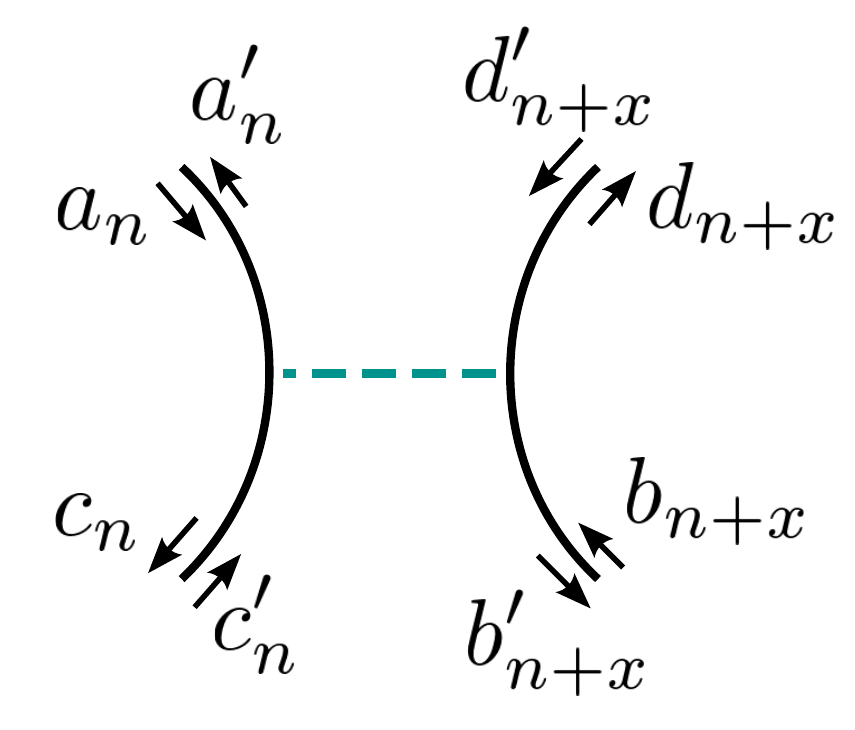}
  \caption{Schematic of inter-resonator couplings, with wave
    amplitudes for both spins included.}
  \label{fig:couplings2}
\end{figure}

It is straightforward to generalize the above calculations to include
both spins, and to include spin-mixing processes.  In
Fig.~\ref{fig:couplings2}, we show the $x$-couplings (the
$y$-couplings are dealt with similarly), which have been augmented to
include the counter-clockwise amplitudes $\{a',b',c',d'\}$.  These
amplitudes, together with the original clockwise counterparts, are
related by the $S$ matrix
\begin{equation}
  \mathbf{S}_x \begin{bmatrix}a_n \\ b_{n+x} \\ d_{n+x}' \\ c_n' \end{bmatrix}
  = \begin{bmatrix}a_n' \\ b_{n+x}' \\ d_{n+x} \\ c_n \end{bmatrix}.
  \label{S4}
\end{equation}
This $4\times4$ $S$ matrix is symmetric, due to optical reciprocity.
If the coupling conserves spin, then the $S$ matrix can be expressed
in terms of the old $2\times 2$ $S$ matrices (\ref{coupling 1}), as
\begin{equation}
  \mathbf{S}_x = \begin{bmatrix}
    \mathbf{0} & S_x^T \\ S_x & \mathbf{0}
  \end{bmatrix}.
\end{equation}
If, on the other hand, there are spin-mixing processes, then the $S$
matrix has a more complicated form, but it remains unitary and
symmetric so long as the system remains reciprocal (i.e.~there are no
nonlinear or magneto-optic materials present).

Suppose the coupling is unitary (no gain or loss) as well as
reciprocal.  If there exists a set of wave amplitudes
$\{a,b,c,d,a',b',c',d'\}$ which satisfies (\ref{S4}), then
\begin{equation}
  \mathbf{S}_x \begin{bmatrix}a_n^{'*} \\ b_{n+x}^{'*} \\ d_{n+x}^*
    \\ c_n^* \end{bmatrix} = \begin{bmatrix}a_n^* \\ b_{n+x}^* \\ d_{n+x}^{'*}
    \\ c_n^{'*} \end{bmatrix}.
  \label{S4}
\end{equation}
This is the time-reversed solution, obtained by replacing $a$ with
$a'^*$, $a'$ with $a^*$, and so on.

The calculation of the projected band structure proceeds along the
same lines as the previous section.  We look for solutions
\begin{equation*}
  [a,b,c,d,a',b',c',d']_{n+x} = e^{iK_x} [a,b,c,d,a',b',c',d']_n.
\end{equation*}
We can easily re-arrange the above $S$ matrices to obtain transfer
matrices, satisfying
\begin{equation}
  \mathbf{M}_j^x \begin{bmatrix}c_j \\ c_j' \\ a_j \\ a_j' \end{bmatrix}
  = e^{iK_x} \, \begin{bmatrix} b_j \\ b_j' \\ d_j \\ d_j'  \end{bmatrix},\quad
  \mathbf{M}^y_j(\phi) \begin{bmatrix}d_j \\ d_j' \\ b_{j+1} \\b_{j+1}' \end{bmatrix}
  = \begin{bmatrix} a_j \\ a_j' \\ c_{j+1} \\c_{j+1}'  \end{bmatrix}.
\end{equation}
The projected band structure is then obtained by solving
\begin{equation}
  \begin{bmatrix} \mathbf{M}_1^x & \\
    & \mathbf{M}_2^x &\\
    & & \ddots\\
  \end{bmatrix} \,
  \begin{bmatrix} e^{-2i\phi} & & & & & & & \\
    & e^{2i\phi} & & & & & & \\
    & & & \mathbf{M}_1^y & & & & \\
    & & & & \ddots & & &\\
    & & & & & \mathbf{M}_{N-1}^y & &\\
    & & & & & & e^{2i\phi} & \\
    & & & & & & & e^{-2i\phi}
  \end{bmatrix}
  \begin{bmatrix} b_1 \\ b_1' \\ d_1 \\ d_1' \\ \vdots \\ d_N \\ d_N'\end{bmatrix}
  = e^{iK_x} \begin{bmatrix} b_1 \\ b_1' \\ d_1 \\ d_1' \\ \vdots \\ d_N\\ d_N' \end{bmatrix}.
  \label{eigenproblem 2}
\end{equation}

In Fig.~\ref{fig:z2}, we show the projected band structure for a
resonator lattice when spin mixing is included.  Here, we model the
inter-resonator couplings by
\begin{equation}
  \mathbf{S}_\mu = e^{i A} \begin{bmatrix} \mathbf{0} &
    S_\mu^T \\ S_\mu & \mathbf{0}
  \end{bmatrix} e^{i A^T},
\end{equation}
where $S_\mu$ is the $2\times 2$ coupling matrix for spin-conserving
couplings (using the same parameters as in Fig.~2 of the paper, with
$\theta = 0.4\pi$), and $A$ is a Hermitian spin-mixing matrix.  The
above form of the coupling matrix is designed to preserve reciprocity
and unitarity.  For illustrative purposes, we pick the spin-mixing
matrix $A$ from the Gaussian unitary ensemble of $4\times 4$ random
Hermitian matrices, multiplied by a scale factor:
\begin{equation}
  A = 0.1 \times \begin{bmatrix}
  -0.4257           &  0.3271 + 0.9193i & -0.4521 + 0.6750i &  0.3534 + 0.4516i \\
   0.3271 - 0.9193i &  0.8429           & -0.1835 + 0.1817i & -0.1301 + 1.7570i \\
  -0.4521 - 0.6750i & -0.1835 - 0.1817i &  1.7557           &  1.1908 - 0.7575i \\
   0.3534 - 0.4516i & -0.1301 - 1.7570i &  1.1908 + 0.7575i &  0.2727
  \end{bmatrix}.
\end{equation}
As in the Kane-Mele model \kanemele, the introduction of spin mixing
lifts the two-fold degeneracy of the edge states.  Each band gap
contains two pairs of edge states; each pair is confined to a single
edge, and consists of two Kramers partners which are related by the
time-reversal operation detailed above.  However, because optical wave
amplitudes are not spin-half objects, the time-reversal operation does
not satisfy $\mathcal{T}^2 = -1$, and Kane and Mele's result
\cite{Kane} that the edge states are protected against
spin-nonconserving perturbations does not hold.  The edge states
therefore receive no protection against spin-mixing perturbations.

\begin{figure}
  \centering\includegraphics[width=7cm]{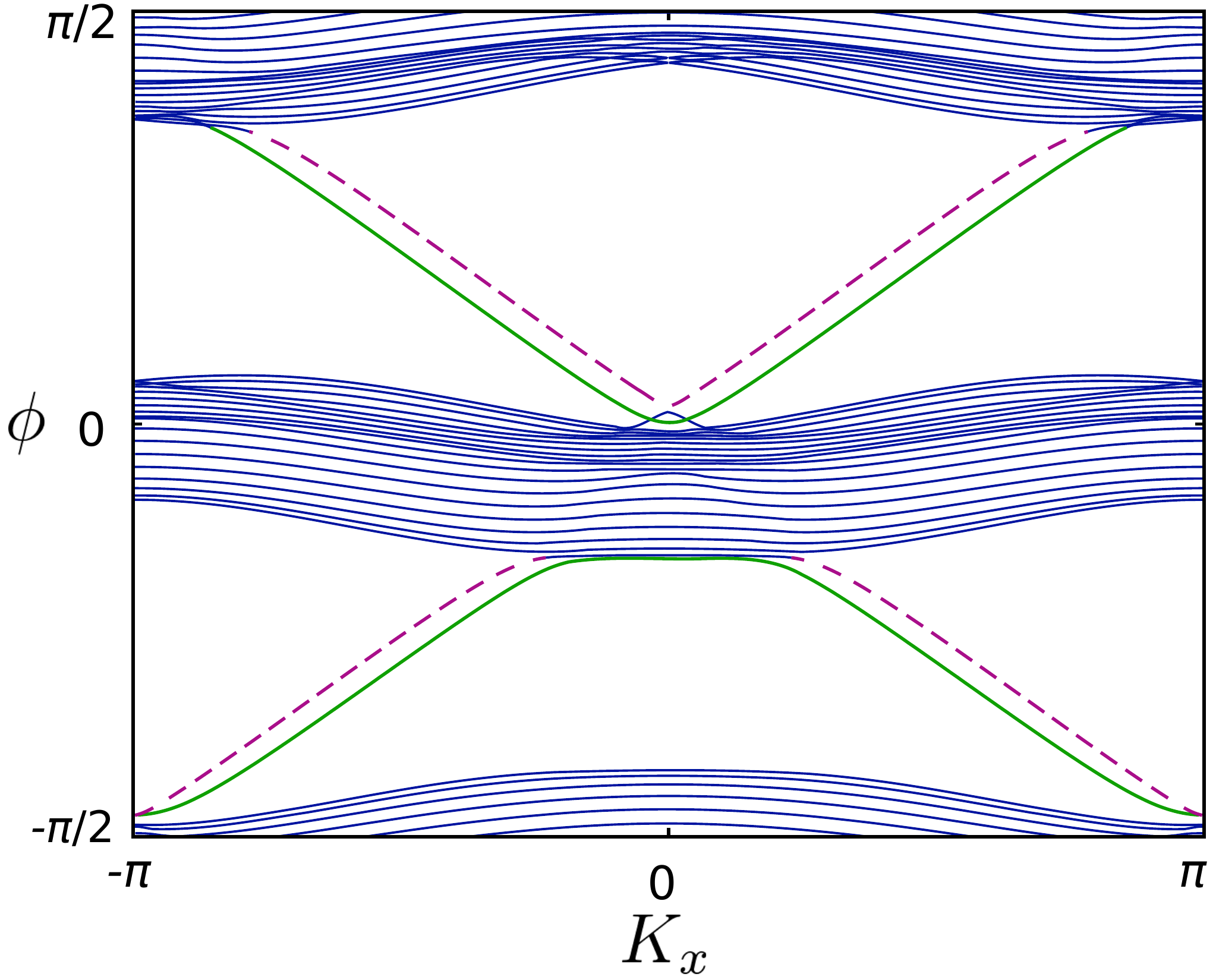}
  \caption{Projected bandstructure with spin-mixing processes.  Solid
    green lines indicate edge states on the lower edge, and dashed
    magenta lines indicate edge states on the upper edge. }
  \label{fig:z2}
\end{figure}

\end{widetext}


\begin{thebibliography}{99}
\bibitem{PT note} This parameterization differs from that of
  Ref.~\cite{LiPT}, which considered symmetric (reciprocal) scattering
  matrices.
\end{thebibliography}
\end{document}